\documentclass[aps,onecolumn,showpacs,nofootinbib,showkeys]{revtex4-2}
\usepackage[margin=3.15cm]{geometry}

\usepackage{graphicx}
\usepackage{changes}
\usepackage{physics}
\usepackage{subcaption}
\usepackage{multirow}
\RequirePackage{hyperref}
\hypersetup{
    linktocpage,
    colorlinks,
    citecolor=blue,
    filecolor=black,
    linkcolor=blue,
    urlcolor=blue,
}
\usepackage{nccmath}

\begin{document}

\title{A theoretical analysis of the doubly radiative decays
$\eta^{(\prime)}\to\pi^0\gamma\gamma$ and $\eta^\prime\to\eta\gamma\gamma$}

\author{Rafel Escribano$^{1,2}$}\email{rescriba@ifae.es}
\author{Sergi Gonz\`{a}lez-Sol\'{i}s$^{3,4,5}$}\email{sgonzal@iu.edu}
\author{Renata Jora$^{6}$}\email{rjora@theory.nipne.ro}
\author{Emilio Royo$^{1,2}$}\email{eroyo@ifae.es}

\affiliation{
$^1$Grup de F\'{i}sica Te\`orica, Departament de F\'{i}sica,
Universitat Aut\`onoma de Barcelona, E-08193 Bellaterra (Barcelona), Spain\\
$^2$Institut de F\'{i}sica d'Altes Energies (IFAE),
The Barcelona Institute of Science and Technology, Campus UAB, 
E-08193 Bellaterra (Barcelona), Spain\\
$^3$CAS Key Laboratory of Theoretical Physics, Institute of Theoretical Physics,
Chinese Academy of Sciences, Beijing 100190, China\\
$^4$Department of Physics, Indiana University, Bloomington, IN 47405, USA\\
$^5$Center for Exploration of Energy and Matter, 
Indiana University, Bloomington, IN 47408, USA\\
$^6$National Institute of Physics and Nuclear Engineering, 
PO Box MG-6, Bucharest-Magurele, Romania
}


\begin{abstract}
The scalar and vector meson exchange contributions to the doubly radiative decays
$\eta^{(\prime)}\to\pi^0\gamma\gamma$ and $\eta^\prime\to\eta\gamma\gamma$
are analysed within the Linear Sigma Model and Vector Meson Dominance frameworks, respectively.
Predictions for the diphoton invariant mass spectra and the associated integrated 
branching ratios are given and compared with current available experimental data.
Whilst a satisfactory description of the shape of the $\eta\to\pi^{0}\gamma\gamma$ and $\eta^\prime\to\pi^{0}\gamma\gamma$ decay spectra is obtained, thus supporting the validity of the approach, the corresponding branching ratios cannot be reproduced simultaneously. 
A first theoretical prediction for the recently measured $\eta^\prime\to\eta\gamma\gamma$ by the BESIII collaboration is also presented.


\end{abstract}


\maketitle


\section{\label{intro}Introduction}
Measurements of $\eta$ and $\eta^{\prime}$ decays have reached 
unprecedented precision over the years placing new demands on the 
accuracy of the corresponding theoretical descriptions \cite{Gan:2020aco}. 
Amongst them, the doubly radiative decay $\eta\to\pi^{0}\gamma\gamma$ 
has attracted much interest as this reaction is a perfect laboratory for testing 
Chiral Perturbation Theory and its natural extensions, but also as a result of 
the decades-long tension between the associated theoretical predictions 
and the experimental measurements.

Likewise, the study of the $\eta^\prime\to\pi^0\gamma\gamma$ and 
$\eta^\prime\to\eta\gamma\gamma$ decay processes are of interest 
for a number of reasons. First, they complete existing calculations of 
the sister process $\eta\to\pi^0\gamma\gamma$, which has been studied 
in many different frameworks, ranging from the seminal works 
based on Vector Meson Dominance (VMD)
\cite{Oppo:1967zz,Baracca:1970bx} and Chiral Perturbation Theory (ChPT) 
\cite{Ametller:1991dp}, to more modern treatments based on the unitarisation 
of the chiral amplitudes \cite{Oset:2002sh,Oset:2008hp} or dispersive 
formalisms \cite{Danilkin:2017lyn}.
At present, whilst there is only a course estimation for the branching 
ratio of the $\eta^{\prime}\to\pi^{0}\gamma\gamma$ decay
\cite{Balytskyi:2018pzb,Balytskyi:2018uxb}, there is no calculation or theoretical 
prediction for the $\eta^\prime\to\eta\gamma\gamma$ branching ratio.
Second, the BESIII collaboration has recently reported the first measurements for
the decays $\eta^{\prime}\to\pi^{0}\gamma\gamma$ \cite{Ablikim:2016tuo}
and $\eta^{\prime}\to\eta\gamma\gamma$ \cite{Ablikim:2019wsb},
thus, making the topic of timely interest \cite{Fang:2017qgz}.
Third, the analysis of these decays could help extract relevant 
information on the properties of the lowest-lying scalar resonances, in particular,
the isovector $a_0(980)$ from the two $\eta^{(\prime)}\to\pi^0\gamma\gamma$ 
processes and the isoscalars $\sigma(500)$ and $f_0(980)$ from the 
$\eta^\prime\to\eta\gamma\gamma$ decay,
thus, complementing other investigations such as the 
studies of $V\to P^0P^0\gamma$ decays
($V=\rho^0, \omega, \phi$ and $P^0=\pi^0, \eta)$ \cite{Escribano:2006mb},
$D$ and $J/\psi$ decays, central production,
etc.~(see note on scalar mesons in Ref.~\cite{PhysRevD.98.030001}).
For all these reasons,
the aim of the present work is to provide a first detailed evaluation
of the invariant mass spectrum and integrated branching ratio for the
three doubly radiative decays $\eta^{(\prime)}\to\pi^0\gamma\gamma$ and
$\eta^\prime\to\eta\gamma\gamma$.
Very preliminary results were presented 
in Refs.~\cite{Escribano:2012dk,Jora:2010zz}.

On the experimental front,
the branching ratio (BR) of the $\eta\to\pi^0\gamma\gamma$ decay 
has been measured by GAMS-2000 \cite{Alde:1984wj},
$\mbox{BR}=(7.1\pm 1.4)\times 10^{-4}$,
CrystalBall@AGS in 2005 \cite{Prakhov:2005vx}, 
$\mbox{BR}=(3.5\pm 0.7\pm 0.6)\times 10^{-4}$, 
and 2008 \cite{Prakhov:2008zz}, $\mbox{BR}=(2.21\pm 0.24\pm 0.47)\times 10^{-4}$,
where the latter also included an invariant mass spectrum for the two outgoing photons.
An independent analysis of the last CrystalBall data resulted in
$\mbox{BR}=(2.7\pm 0.9\pm 0.5)\times 10^{-4}$ \cite{Knecht:2004gx}.
Early results are summarised in the review of Ref.~\cite{Landsberg:1986fd}.
Surprisingly low in comparison with all previous measurements is
the 2006 result reported by the KLOE collaboration \cite{DiMicco:2005stk},
$\mbox{BR}=(0.84\pm 0.27\pm 0.14)\times 10^{-4}$,
based on a sample of $68\pm 23$ events.
More recently, a new measurement of the diphoton energy spectrum,
as well as new and more precise values for 
$\Gamma(\eta\to\pi^{0}\gamma\gamma)=(0.330\pm 0.030)$ eV and
$\mbox{BR}=(2.54\pm 0.27)\times 10^{-4}$ have been released by the A2 collaboration at the Mainz Microtron (MAMI), based on the analysis of 
$1.2\times 10^{3}$ $\eta\to\pi^{0}\gamma\gamma$ decay events
\cite{Nefkens:2014zlt}.
The latest PDG states a fit value of $\mbox{BR}=(2.56\pm 0.22)\times 10^{-4}$
\cite{PhysRevD.98.030001}.
For the $\eta^\prime\to\pi^0\gamma\gamma$ decay,
the BESIII collaboration has recently reported for the first time
the associated $m_{\gamma\gamma}^{2}$ invariant mass distribution
\cite{Ablikim:2016tuo}.
The measured branching fraction is 
$\mbox{BR}=(3.20\pm 0.07\pm 0.23)\times10^{-3}$,
superseding the upper limit
$\mbox{BR}<8 \times 10^{-4}$ at 90\% CL
determined by the GAMS-2000 experiment \cite{Alde:1987jt}.
Finally, for the $\eta^\prime\to\eta\gamma\gamma$ decay, 
a measurement of $\mbox{BR}<1.33\times 10^{-4}$ at 90\% CL 
has been provided, again for the first time, by the BESIII collaboration \cite{Ablikim:2019wsb}.

On the theoretical front,
the $\eta\to\pi^0\gamma\gamma$ decay has been a stringent test for the
predictive power of ChPT. Within this framework,
the tree-level contributions at ${\cal O}(p^2)$ and ${\cal O}(p^4)$
vanish because the pseudoscalar mesons involved are neutral.
The first non-vanishing contribution comes at ${\cal O}(p^4)$,
either from kaon loops, largely suppressed by their mass, 
or from pion loops, also suppressed since they violate $G$-parity
and, therefore, are proportional to $m_u-m_d$.
Quantitatively, Ametller \textit{et al.}~found in Ref.~\cite{Ametller:1991dp} 
that $\Gamma_\pi^{(4)}=0.84\times 10^{-3}$ eV,
$\Gamma_K^{(4)}=2.45\times 10^{-3}$ eV and
$\Gamma_{\pi,\, K}^{(4)}=3.89\times 10^{-3}$ eV 
for the $\pi$, $K$ and $\pi$+$K$ loop contributions 
to the decay width, which turns out to be two orders of magnitude 
smaller than the PDG fit value
$\Gamma_{\eta\to\pi^0\gamma\gamma}^{\mathrm{exp}}=0.334\pm 0.028$ eV
\cite{PhysRevD.98.030001}.
The first sizeable contribution comes at ${\cal O}(p^6)$,
but the associated low-energy constants are not well defined and one must 
resort to phenomenological models to fix them.
To this end, for instance, VMD has been used to determine these coefficients
by expanding the vector meson propagators and keeping the lowest term.
Assuming equal contributions from the $\rho^0$ and $\omega$ mesons,
the authors of Ref.~\cite{Ametller:1991dp} found that
$\Gamma_{\rho+\omega}^{(6)}=0.18$ eV, which was
about two times smaller than their ``all-order'' estimation
with the full vector meson propagator $\Gamma_{\mathrm{VMD}}=0.31$ eV, 
and in reasonable agreement with older VMD estimates 
\cite{Oppo:1967zz,Baracca:1970bx},
as well as Refs.~\cite{Picciotto:1991sn,Ng:1992yg}. 
The contributions of the scalar $a_0(980)$ and
tensor $a_2(1320)$ resonances to the ${\cal O}(p^6)$ chiral coefficients were 
also assessed in Ref.~\cite{Ametller:1991dp} following the same procedure
but no ``all-order'' estimates were provided.
Contrary to the VMD contribution where the coupling constants
appear squared, the signs of the $a_0$ and $a_2$ contributions are not
unambiguously fixed \cite{Ametller:1991dp}.
At order ${\cal O}(p^8)$, a new type of loop effects taking two vertices
from the anomalous chiral Lagrangian appear.
Pion loops are no longer suppressed since the associated vertices do not
violate $G$-parity and the kaon-loop suppression does not necessarily occur.
Numerically, the contributions from these loops were \cite{Ametller:1991dp}
$\Gamma_\pi^{(8)}=5.2\times 10^{-5}$ eV,
$\Gamma_K^{(8)}=2.2\times 10^{-3}$ eV and
$\Gamma_{\pi,K}^{(8)}=2.5\times 10^{-3}$ eV.
Summing up all the effects that were not negligible and
presented no sign ambiguities,
\textit{i.e.}~the non-anomalous pion and kaon loops at ${\cal O}(p^4)$,
the corresponding loops at ${\cal O}(p^8)$ with two anomalous vertices,
and the ``all-order'' VMD estimate, resulted in 
$\Gamma_{\eta\to\pi^0\gamma\gamma}^{\chi+\mathrm{VMD}}=0.42$ eV \cite{Ametller:1991dp}.
Including the contributions from the $a_0$ and $a_2$ 
exchanges with sign ambiguities,
which did not represent an ``all-order'' estimate of these effects,
they conservatively concluded that 
$\Gamma_{\eta\to\pi^0\gamma\gamma}^{\chi+\mathrm{VMD}+a_0+a_2}=
0.42\pm 0.20$ eV \cite{Ametller:1991dp}.
The further inclusion of $C$-odd axial-vector resonances raised this value to 
$0.47\pm 0.20$ eV \cite{Ko:1992zr} (see also Ref.~\cite{Ko:1993rg}).
Other determinations of the ${\cal O}(p^6)$ low-energy constants
in the early and extended Nambu--Jona-Lasinio models led to 0.11--0.35 eV
\cite{Belkov:1995qdq}, $0.58\pm 0.30$ eV \cite{Bellucci:1995ay}
and $0.27^{+0.18}_{-0.07}$ eV \cite{Bijnens:1995vg}.
A different approach based on quark-box diagrams \cite{Ng:1993sc,Nemoto:1996bh}
yielded values of 0.70 eV and 0.58--0.92 eV, respectively.
In the most recent analyses, the $\eta\to\pi^0\gamma\gamma$ process
has been considered within a chiral unitary approach for the meson-meson 
interaction, thus generating the $a_0$ resonance and fixing 
the sign ambiguity of its contribution.
Using this approach, Oset \textit{et al.}~found a decay width of 
$0.47\pm 0.10$ eV and $0.33\pm 0.08$ eV
in their 2003 \cite{Oset:2002sh} and 2008 \cite{Oset:2008hp} works, respectively,
and the discrepancy could be down to differences in the radiative decay 
widths of the vector mesons used as input in their calculations. 
In any case, both estimations appear to be in good 
agreement with the empirical value
$\Gamma_{\eta\to\pi^0\gamma\gamma}^{\mathrm{exp}}=0.334\pm 0.028$ eV.
On the other hand, there is only a rough estimation for the
$\eta^\prime\to\pi^0\gamma\gamma$ decay width \cite{Balytskyi:2018pzb,Balytskyi:2018uxb}
and no theoretical analysis for the $\eta^\prime\to\eta\gamma\gamma$ process.

The methodology in the present work can be summarised as follows.
First, we begin calculating the dominant chiral-loop contribution,
that is, the ${\cal O}(p^4)$ diagrams containing two vertices of the lowest order 
Lagrangian and one loop of charged pions or kaons. We employ 
the large-$N_c$ limit of ChPT 
and regard the singlet state $\eta_0$ as the ninth 
pseudo-Goldstone boson of the theory.
In addition, we simplify the calculations by assuming 
the isospin limit, which allows one to consider only the kaon loops
for the two $\eta^{(\prime)}\to\pi^0\gamma\gamma$ decays.
The ${\cal O}(p^8)$ loop corrections from diagrams with two anomalous vertices 
are very small \cite{Ametller:1991dp} and, therefore, not considered.
The explicit contributions of intermediate vector and scalar mesons are accounted for
by means of the VMD and Linear Sigma Model (L$\sigma$M) frameworks.
Accordingly, we compute the dominant contribution, \textit{i.e.}~the 
exchange of intermediate vector mesons, through the decay chain
$P^0\to V\gamma\to P^0\gamma\gamma$. 
Next, we consider the scalar meson contributions,
providing an ``all-order'' estimate of the scalar effects,
through a calculation performed within the L$\sigma$M,
which enables us to, first, fix the sign ambiguity and, second, 
assess the relevance of the full scalar meson propagators,
as opposed to integrating them out. 

The structure of this work is as follows.
In section \ref{chpt}, we review the ChPT calculation for the
$\eta\to\pi^{0}\gamma\gamma$ and provide theoretical expressions for the 
$\eta^{\prime}\to\pi^{0}\gamma\gamma$ and 
$\eta^{\prime}\to\eta\gamma\gamma$ decays.
In section \ref{VMD}, we calculate the effects of intermediate 
vector meson exchanges, which represent the dominant contribution,
using the VMD model. 
In section \ref{LsM}, the chiral-loop prediction is substituted by a L$\sigma$M
calculation where the effects of scalar meson resonances 
are taken into account explicitly.
In section \ref{results}, theoretical results for the decay widths and 
associated diphoton energy spectra are presented for the three decay processes
$\eta^{(\prime)}\to\pi^0\gamma\gamma$ and $\eta^\prime\to\eta\gamma\gamma$,
and a detailed discussion of the results is given.
Some final remarks and conclusions are presented in section \ref{Conclusions}.



\section{\label{chpt}Chiral-loop calculation}

Let us focus our attention to the $\eta\to\pi^0\gamma\gamma$ process.
At order ${\cal O}(p^2)$, there are no contributions to this 
process and, at ${\cal O}(p^4)$, the contributions come from 
diagrams with two vertices from the lowest order chiral Lagrangial and
a loop of charged pions and kaons. However, as discussed 
in section \ref{intro}, the contribution 
from kaon loops is dominant and the pion loops vanish in the isospin limit.
The invariant amplitude can, thus, be written as follows 
\begin{equation}
\label{Achietapi0}
{\cal A}^{\mathrm{\chi PT}}_{\eta\to\pi^0\gamma\gamma}=
\frac{2\alpha}{\pi}\frac{1}{m_{K^+}^2}L(s_K)\{a\}\times{\cal A}^\chi_{K^+K^-\to\pi^0\eta}\ ,
\end{equation}
where
$\alpha$ is the fine-structure constant, $m_{K^+}$ is the mass of the 
charged kaon, $L(\hat s)$ is the loop integral 
\begin{equation}
\begin{aligned}
\label{L}
L(z)&=-\frac{1}{2z}-\frac{2}{z^2}f\left(\frac{1}{z}\right)\ ,\\[1ex]
f(z)&=\left\{
\begin{array}{ll}
\frac{1}{4}\left(\log\frac{1+\sqrt{1-4z}}{1-\sqrt{1-4z}}-i\pi\right)^2 & \mbox{for}\ z<\frac{1}{4}\\[1ex]
-\left[\arcsin\left(\frac{1}{2\sqrt{z}}\right)\right]^2 & \mbox{for}\ z>\frac{1}{4}
\end{array}\right.\, ,
\end{aligned}
\end{equation}
and $s_K=s/m_{K^+}^2$, with $s=(q_1+q_2)^2=2q_1\cdot q_2$
being the invariant mass of the two outgoing photons. 
The Lorentz structure $\{a\}$ in Eq.~\eqref{Achietapi0} is defined as
\begin{equation}
\{a\}=(\epsilon_1\cdot\epsilon_2)(q_1\cdot q_2)-(\epsilon_1\cdot q_2)(\epsilon_2\cdot q_1) \ ,
\label{astruct}
\end{equation}
where $\epsilon_{1,2}$ and $q_{1,2}$ are 
the polarisation and four-momentum vectors of the final photons and
${\cal A}^\chi_{K^+K^-\to\pi^0\eta}$ is the four-pseudoscalar amplitude, 
which can be expressed as follows\footnote{This amplitude should not be confused with the four-pseudoscalar scattering amplitude calculated in ChPT at lowest order.}
\begin{widetext}
\begin{equation}
\begin{aligned}
\label{AChPTKpKpi0eta}
{\cal A}^{\mathrm{\chi PT}}_{K^+K^-\to\pi^0\eta}=&\frac{1}{4f_\pi^2}
\bigg[\bigg(s-\frac{m_\eta^2}{3}-\frac{8m_K^2}{9}-\frac{m_\pi^2}{9}\bigg)(\cos\varphi_P+\sqrt{2}\sin\varphi_P) \\
&+\frac{4}{9}(2m_K^2+m_\pi^2)\bigg(\cos\varphi_P-\frac{\sin\varphi_P}{\sqrt{2}}\bigg)\bigg]\, ,
\end{aligned}
\end{equation}
\end{widetext}
where $f_{\pi}$ is the pion decay constant and $\varphi_P$ is the 
$\eta$-$\eta^{\prime}$ pseudoscalar mixing angle in the quark-flavour basis
at lowest order in ChPT defined as
\begin{equation}
\begin{aligned}
\ket{\eta}=\cos{\varphi_P}\ket{\eta_{\textrm{NS}}}-\sin{\varphi_P}\ket{\eta_{\textrm{S}}}\ ,\\
\ket{\eta^{\prime}}=\sin{\varphi_P}\ket{\eta_{\textrm{NS}}}+\cos{\varphi_P}\ket{\eta_{\textrm{S}}}\ , 
\end{aligned}
\label{eq101}
\end{equation}
with $\ket{\eta_{\textrm{NS}}} = \frac{1}{\sqrt{2}}\ket{u\bar{u} + d\bar{d}}$ and 
$\ket{\eta_{\textrm{S}}} = \ket{s\bar{s}}$ \cite{Bramon:1997mf}.

It must be noted that, in the seminal work of Ref.~\cite{Ametller:1991dp},
the chiral-loop prediction was computed taking only into account the $\eta_8$ 
contribution and the mixing angle was fixed to
$\theta_P=\varphi_P-\arctan\sqrt{2}=\arcsin(-1/3)\simeq -19.5^\circ$.
As explained before, in this work the singlet contribution is also considered 
and the dependence on the mixing angle is made explicit.

For the $\eta^\prime\to\pi^0\gamma\gamma$ process,
the associated amplitude is that of Eq.~(\ref{Achietapi0}) with the 
replacements $m_\eta\to m_{\eta^\prime}$,
$(\cos\varphi_P+\sqrt{2}\sin\varphi_P)\to (\sin\varphi_P-\sqrt{2}\cos\varphi_P)$ and
$(\cos\varphi_P-\sin\varphi_P/\sqrt{2})\to (\sin\varphi_P+\cos\varphi_P/\sqrt{2})$ in Eq.~(\ref{AChPTKpKpi0eta}).
Finally, for the $\eta^\prime\to\eta\gamma\gamma$ decay, 
two types of amplitudes contribute,
one associated to a loop of charged kaons, as in the former two cases, 
and the other to a loop of charged pions, which in this case is not 
suppressed by $G$-parity. Again, the corresponding amplitudes have the 
same structure as Eq.~(\ref{Achietapi0}) but replacing $s_K\to s_\pi$ 
and $m_{K^+}\to m_{\pi^+}$ for the pion loop, and, 
instead of Eq.~(\ref{AChPTKpKpi0eta}), one must make use of
\begin{widetext}
\begin{fleqn}
\begin{equation}
\label{AChPTKpKetaetap}
\begin{aligned}
\qquad\qquad{\cal A}^{\mathrm{\chi PT}}_{K^+K^-\to\eta\eta^\prime}=&-\frac{1}{4f_\pi^2}
\bigg[\bigg(s-\frac{m_\eta^2+m_{\eta^\prime}^2}{3}-\frac{8m_K^2}{9}-\frac{2m_\pi^2}{9}\bigg)
\left(\sqrt{2}\cos2\varphi_P+\frac{\sin2\varphi_P}{2}\right) \\[2pt]
&+\frac{4}{9}(2m_K^2-m_\pi^2)\left(2\sin2\varphi_P-\frac{\cos2\varphi_P}{\sqrt{2}}\right)\bigg]\ , \\[-1ex]
\end{aligned}
\end{equation}
\begin{equation}
\label{AChPTpippimetaetap}
\begin{aligned}
\qquad\qquad{\cal A}^{\mathrm{\chi PT}}_{\pi^+\pi^-\to\eta\eta^\prime}=\frac{m_\pi^2}{2f_\pi^2}\sin2\varphi_P\ ,
\end{aligned}
\end{equation}
\end{fleqn}
\end{widetext}
for the loop of kaons and pions, respectively.

To the best of our knowledge, the amplitudes for the 
$\eta^\prime\to\pi^0\gamma\gamma$ and $\eta^\prime\to\eta\gamma\gamma$
constitute the first chiral-loop predictions for these processes.
\\


\section{\label{VMD}VMD calculation}
As discussed in section \ref{intro}, VMD can be used to calculate an 
``all-order'' estimate for the contribution of intermediate vector meson 
exchanges to the processes of interest in this work.
In Ref.~\cite{Ametller:1991dp}, 
for example, it was found that the VMD amplitude represents
the dominant contribution to the $\eta\to\pi^0\gamma\gamma$ decay,
and, as it will be shown in Section \ref{results}, this is also the case
for the $\eta^\prime\to\pi^0\gamma\gamma$ and 
$\eta^\prime\to\eta\gamma\gamma$ processes.


There is a total of six Feynman diagrams contributing to each one 
of the three decay processes, corresponding to the exchange of the 
three neutral vector mesons $\rho^0$, $\omega$ and $\phi$.
After some algebra, one arrives at the following expression for 
the invariant amplitude of the $\eta\to\pi^0\gamma\gamma$ decay
\begin{widetext}
\begin{eqnarray}
\label{AVMDetapi0}
\quad {\cal A}^{\mathrm{VMD}}_{\eta\to\pi^0\gamma\gamma}=
\sum_{V=\rho^0, \omega, \phi}g_{V\!\eta\gamma}g_{V\!\pi^0\gamma}\left[\frac{(P\cdot q_2-m_\eta^2)\{a\}-\{b\}}{D_V(t)}+
\bigg\{
\begin{array}{c}
q_2\leftrightarrow q_1\\
t\leftrightarrow u
\end{array}
\bigg\}\right]\ ,
\end{eqnarray}
\end{widetext}
where
$t,u=(P-q_{2,1})^2=m_\eta^2-2P\cdot q_{2,1}$ are the Mandelstam variables,
and the Lorentz structures $\{a\}$ and $\{b\}$ are defined as
\begin{equation}
\label{varGamma}
\begin{aligned}
\{a\}&=(\epsilon_1\cdot\epsilon_2)(q_1\cdot q_2)-(\epsilon_1\cdot q_2)(\epsilon_2\cdot q_1) \ , \\
\{b\}&=(\epsilon_1\cdot q_2)(\epsilon_2\cdot P)(P\cdot q_1)+(\epsilon_2\cdot q_1)(\epsilon_1\cdot P)(P\cdot q_2)-(\epsilon_1\cdot\epsilon_2)(P\cdot q_1)(P\cdot q_2)\\
&-(\epsilon_1\cdot P)(\epsilon_2\cdot P)(q_1\cdot q_2) \ ,
\end{aligned}
\end{equation}
where $P$ is the four-momentum of the decaying particle, and
$\epsilon_{1,2}$ and $q_{1,2}$ are the polarisation and four-momentum 
vectors of the final photons, respectively. The denominator
$D_V(t)=m_V^2-t-i\,m_V\Gamma_V$ is the vector meson propagator,
with $V=\omega$ and $\phi$; for the $\rho^0$ propagator, we use, 
instead, an energy-dependent decay width
\begin{equation}
\label{varGamma}
\Gamma_{\rho^0}(t)=\Gamma_{\rho^0}\times[(t-4m_\pi^2)/(m_{\rho^0}^2-4m_\pi^2)]^{3/2}\times\theta(t-4m_\pi^2) \ .
\end{equation}
The amplitudes for the $\eta^\prime\to\pi^0\gamma\gamma$ 
and $\eta^\prime\to\eta\gamma\gamma$ decays have a similar 
structure to that of Eq.~(\ref{AVMDetapi0}), with the replacements
$m_\eta^2\to m_{\eta^\prime}^2$, and $g_{V\eta\gamma}g_{V\pi^0\gamma}\to 
g_{V\eta^\prime\gamma}g_{V\pi^0\gamma}$ for the $\eta^\prime\to\pi^0\gamma\gamma$ and
$g_{V\eta\gamma}g_{V\pi^0\gamma}\to g_{V\eta^\prime\gamma}g_{V\eta\gamma}$
for the $\eta^\prime\to\eta\gamma\gamma$ case.

To parametrise the $V\!P\gamma$ coupling constants, $g_{V\!P\gamma}$, 
one can make use of a simple phenomenological quark-based model 
first presented in Ref.~\cite{Bramon:2000fr}, which was initially developed 
to describe $V\to P\gamma$ and $P\to V\gamma$ radiative decays. 
The coupling constants can, thus, be written as
\cite{Bramon:2000fr,Escribano:2020jdy}
\begin{equation}
\begin{aligned}
g_{\rho^0\pi^0\gamma} &= \frac{1}{3}g\ \!, \\
g_{\rho^0\eta\gamma} &= gz_{\textrm{NS}}\cos{\varphi_P}\ \!, \\ 
g_{\rho^0\eta^{\prime}\gamma} &= gz_{\textrm{NS}}\sin{\varphi_P}\ \!, \\
g_{\omega \pi^0 \gamma} &= g\cos{\varphi_V}\ \!, \\
g_{\omega\eta\gamma} &= \frac{1}{3}g\Big(z_{\textrm{NS}}\cos{\varphi_P}\cos{\varphi_V} - 2 \frac{\overline{m}}{m_s}z_{\textrm{S}}\sin{\varphi_P}\sin{\varphi_V}\Big)\ \!, \\
g_{\omega\eta^{\prime}\gamma} &= \frac{1}{3}g\Big(z_{\textrm{NS}}\sin{\varphi_P}\cos{\varphi_V} + 2 \frac{\overline{m}}{m_s}z_{\textrm{S}}\cos{\varphi_P}\sin{\varphi_V}\Big)\ \!, \\
g_{\phi\pi^0\gamma} &= g\sin{\varphi_V}\ \!, \\
g_{\phi\eta\gamma} &= \frac{1}{3}g\Big(z_{\textrm{NS}}\cos{\varphi_P}\sin{\varphi_V} + 2 \frac{\overline{m}}{m_s}z_{\textrm{S}}\sin{\varphi_P}\cos{\varphi_V}\Big)\ \!, \\
g_{\phi\eta^{\prime}\gamma} &= \frac{1}{3}g\Big(z_{\textrm{NS}}\sin{\varphi_P}\sin{\varphi_V} - 2 \frac{\overline{m}}{m_s}z_{\textrm{S}}\cos{\varphi_P}\cos{\varphi_V}\Big)\ \!,
\end{aligned}
\label{eqcoups}
\end{equation}
where $g$ is a generic electromagnetic constant, 
$\varphi_P$ is, again, the pseudoscalar $\eta$-$\eta^{\prime}$ mixing angle 
in the quark-flavour basis, $\varphi_V$ is the vector $\omega$-$\phi$ 
mixing angle in the same basis, 
$\overline{m}/m_s$ is the quotient of constituent quark masses\footnote{The
flavour symmetry-breaking mechanism associated to 
differences in the effective magnetic moments of light (\textit{i.e.}~up and down) 
and strange quarks in magnetic dipolar transitions is implemented 
via constituent quark mass differences. Specifically, one introduces 
a multiplicative $SU(3)$-breaking term, \textit{i.e.}~$1-s_e \equiv \overline{m}/m_s$, 
in the $s$-quark entry of the quark-charge matrix $Q$.}, 
and $z_{\textrm{NS}}$ and $z_{\textrm{S}}$ are the \textit{non-strange} 
and \textit{strange} multiplicative factors accounting for the relative meson 
wavefunction overlaps \cite{Bramon:2000fr,Escribano:2020jdy}. 

It is important to note that in Ref.~\cite{Ametller:1991dp}, the VMD prediction
for the $\eta\to\pi^0\gamma\gamma$ process
was calculated assuming equal $\rho^0$ and $\omega$ contributions 
and without including the decay widths in the propagators.
These approximations were valid in this particular case, 
since the phase space available 
prevents the vector mesons to resonate.
However, for the $\eta^\prime\to\pi^0\gamma\gamma$ case,
the available phase space allows these vectors to be on-shell 
and the introduction of their decay widths 
is mandatory. For consistency, we include the decay widths in the 
vector meson propagators of the three decays of interest in this work.


\section{\label{LsM}L$\sigma$M calculation}
An ``all-order" estimate for the contribution of scalar meson exchanges 
to the processes under study can be obtained by means of the L$\sigma$M,
where the complementarity between this model and ChPT can be used 
to include the scalar meson poles at the same time as keeping the correct 
low-energy behaviour expected from chiral symmetry.
This procedure 
was applied with success to the related $V\to P^0 P^0\gamma$ decays 
\cite{Escribano:2006mb}.

Within this framework, the two $\eta^{(\prime)}\rightarrow\pi^{0}\gamma\gamma$ 
processes proceed through kaon loops and by exchanging the $a_0(980)$ 
in the $s$-channel and the $\kappa$ in the $t$- and $u$-channels.
The $\eta^{\prime}\rightarrow\eta\gamma\gamma$ decay is more complex, 
as it proceeds through both kaon and pion loops; the $\sigma(600)$ 
and the $f_0(980)$ are exchanged in the $s$-channel for both types of loops, 
whilst, in the $u$- and $t$-channels, the $\kappa$ is exchanged for 
kaon loops and the $a_{0}(980)$ for pion loops.

The loop contributions take place through combinations of three diagrams 
for each one of the intermediate states, which added together give finite results.
The amplitudes for the three $\eta^{(\prime)}\rightarrow\pi^0\gamma\gamma$ 
and $\eta^{\prime}\rightarrow\eta\gamma\gamma$
processes in the L$\sigma$M can, 
thus, be expressed as follows 
\begin{widetext}
\begin{eqnarray}
{\cal A}^{\mathrm{L\sigma M}}_{\eta\to\pi^0\gamma\gamma}&=&
\frac{2\alpha}{\pi}\frac{1}{m_{K^+}^2}L(s_K)\{a\}\times{\cal A}^{\rm{L\sigma M}}_{K^+K^-\to\pi^0\eta}\,,\label{Aetapi0LsM}\\[1ex]
{\cal A}^{\mathrm{L\sigma M}}_{\eta^{\prime}\to\pi^0\gamma\gamma}&=&
\frac{2\alpha}{\pi}\frac{1}{m_{K^+}^2}L(s_K)\{a\}\times{\cal A}^{\rm{L\sigma M}}_{K^+K^-\to\pi^0\eta^{\prime}}\,,\label{Aetappi0LsM}\\[1ex]
{\cal A}^{\mathrm{L\sigma M}}_{\eta^{\prime}\to\eta\gamma\gamma}&=&
\frac{2\alpha}{\pi}\frac{1}{m_{\pi}^2}L(s_\pi)\{a\}\times{\cal A}^{\rm{L\sigma M}}_{\pi^+\pi^-\to\eta\eta^{\prime}}+\frac{2\alpha}{\pi}\frac{1}{m_{K^+}^2}L(s_K)\{a\}\times{\cal A}^{\rm{L\sigma M}}_{K^+K^-\to\eta\eta^{\prime}}\,,\label{AetapetaLsM}
\end{eqnarray}
\end{widetext}
where $L(z)$, $s_{\pi,K}$ and $\{a\}$ are the same as in section \ref{chpt}.
The four-pseudoscalar amplitudes 
${\cal A}_{\eta^{(\prime)}\pi^{0}\rightarrow K^+K^-}^{\rm{L\sigma M}}$ 
and ${\cal A}_{\eta^{\prime}\eta\rightarrow K^+K^-(\pi^+\pi^-)}^{\rm{L\sigma M}}$ 
in Eqs.~\eqref{Aetapi0LsM}-\eqref{AetapetaLsM} 
turn out to be $s$, $t$ and $u$ dependent 
and can be expressed in terms of the pion and kaon decay constants, 
$f_{\pi}$ and $f_K$, the masses of the scalar and pseudoscalar 
mesons involved in the processes, and the scalar and pseudoscalar 
mixing angles in the quark-flavour basis, $\varphi_{S}$ and $\varphi_{P}$, 
where $\varphi_S$ is defined as
\begin{equation}
\begin{aligned}
\ket{\sigma}&=\cos\varphi_S\ket{\sigma_{\rm{NS}}}-\sin\varphi_S\ket{\sigma_{\rm{S}}}\ , \\[4pt]
\ket{f_0}&=\sin\varphi_S\ket{\sigma_{\rm{NS}}}+\cos\varphi_S\ket{\sigma_{\rm{S}}} ,
\end{aligned}
\end{equation}
with $\ket{\sigma_{\rm{NS}}}=\frac{1}{\sqrt{2}}\ket{u\bar u+d\bar d}$ 
and $\ket{\sigma_{\rm{S}}}= \ket{s\bar s}$. 
For our analysis, the procedure outlined in Ref.~\cite{Escribano:2006mb} 
is applied in order to obtain a consistent full $s$-dependent amplitude. 
In essence, this involves replacing the $t$- and $u$-channel contributions 
by the result of subtracting from the chiral-loop amplitude, 
\textit{i.e.}~Eqs.~\eqref{AChPTKpKpi0eta}-\eqref{AChPTpippimetaetap}, 
the infinite mass limit of the $s$-channel 
scalar contribution\footnote{It is important to note that this approximation 
is possible due to the fact that, in the $t$- and $u$-channels, the exchanged 
scalar mesons do not resonate. 
}.
We refer the interested reader to Ref.~\cite{Escribano:2006mb} for 
further details. After performing these replacements, one finally 
obtains the following scalar amplitudes
\\
\begin{widetext}
\begin{fleqn}
\begin{equation}
\begin{aligned}
\label{AKpKmpi0etaChPTLsM}
{\cal A}_{K^+K^-\rightarrow\pi^0\eta}^{\mbox{\scriptsize L$\sigma$M}}
=&\frac{1}{2f_\pi f_K}\bigg\{(s-m^2_\eta)\frac{m^2_K-m^2_{a_0}}{D_{a_{0}}(s)}\cos\varphi_P+\frac{1}{6}\Big[(5m_{\eta}^{2}+m_{\pi}^{2}-3s)\cos\varphi_{P} \\
&-\sqrt{2}(m_{\eta}^{2}+4m_{K}^{2}+m_{\pi}^{2}-3s)\sin\varphi_{P}\Big]\bigg\}\ ,
\end{aligned}
\end{equation}
\begin{equation}
\label{Apipimpi0etapChPTLsM}
\begin{aligned}
{\cal A}_{K^+K^-\rightarrow\pi^0\eta^{\prime}}^{\mbox{\scriptsize L$\sigma$M}}
=&\frac{1}{2f_\pi f_K}\bigg\{(s-m^2_{\eta^{\prime}})\frac{m^2_K-m^2_{a_0}}{D_{a_{0}}(s)}\sin\varphi_P+\frac{1}{6}\Big[(5m_{\eta^{\prime}}^{2}+m_{\pi}^{2}-3s)\sin\varphi_{P} \\
&+\sqrt{2}(m_{\eta^{\prime}}^{2}+4m_{K}^{2}+m_{\pi}^{2}-3s)\cos\varphi_{P}\Big]\bigg\}\ ,
\end{aligned}
\end{equation}
\begin{equation}
\label{AKpKmetaetapChPTLsM}
\begin{aligned}
{\cal A}_{K+K^-\rightarrow\eta\eta^{\prime}}^{\mbox{\scriptsize L$\sigma$M}}
=&\frac{s-m_{K}^{2}}{2f_{K}}\bigg[\frac{g_{\sigma\eta\eta^{\prime}}}{D_{\sigma}(s)}\big(\cos\varphi_{S}-\sqrt{2}\sin\varphi_{S}\big)
+\frac{g_{f_{0}\eta\eta^{\prime}}}{D_{f_{0}}(s)}\big(\sin\varphi_{S}+\sqrt{2}\cos\varphi_{S}\big)\bigg]\\[1ex]
&-\frac{s-m_{K}^{2}}{4f_{\pi}f_{K}}\left[1-2\bigg(\frac{2f_{K}}{f_{\pi}}-1\bigg)\right]\sin(2\varphi_{P}) 
-\frac{1}{4f_\pi^2}
\bigg[\bigg(s-\frac{m_\eta^2+m_{\eta^\prime}^2}{3}-\frac{8m_K^2}{9}-\frac{2m_\pi^2}{9}\bigg) \\[1ex]
&\times\left(\sqrt{2}\cos2\varphi_P+\frac{\sin2\varphi_P}{2}\right)+\frac{4}{9}(2m_K^2-m_\pi^2)\left(2\sin2\varphi_P-\frac{\cos2\varphi_P}{\sqrt{2}}\right)\bigg]\ ,
\end{aligned}
\end{equation}
\begin{equation}
\label{ApippimetaetapChPTLsM}
\begin{aligned}
{\cal A}_{\pi^+\pi^-\rightarrow\eta\eta^{\prime}}^{\mbox{\scriptsize L$\sigma$M}}
=\frac{s-m^2_\pi}{f_\pi}\bigg[\frac{g_{\sigma\eta\eta^\prime}}{D_{\sigma}(s)}\cos\varphi_{S}
+\frac{g_{f_0\eta\eta^\prime}}{D_{f_{0}}(s)}\sin\varphi_S
\bigg]+\frac{2m_{\pi}^{2}-s}{2f_{\pi}^2}\sin 2\varphi_{P}\ ,
\end{aligned}
\end{equation}
\end{fleqn}
\end{widetext}
where $D_{S}(s)$ are the $S=\sigma,f_{0}$ and $a_{0}$ propagators 
defined in Appendix~\ref{propagators}.
Note that they are complete one-loop propagators, as the usual 
Breit-Wigner description is not adequate in this case due to either 
the presence of thresholds or a very wide decay width.
The required couplings in  Eqs.~(\ref{AKpKmetaetapChPTLsM}) and
(\ref{ApippimetaetapChPTLsM}) are given by
\begin{widetext}
\begin{fleqn}
\begin{equation}
\label{gsigmaetaetap}
\begin{aligned}
\qquad\quad g_{\sigma\eta\eta^\prime}=&\frac{\sin2\varphi_P}{2f_\pi}
\bigg\{(m^2_\eta\cos^2\varphi_P+m^2_{\eta^\prime}\sin^2\varphi_P-m^2_{a_0})
\left[\cos\varphi_S+\sqrt{2}\sin\varphi_S\left(2\frac{f_K}{f_\pi}-1\right)\right] \\[1ex]
&-(m^2_{\eta^\prime}-m^2_\eta)\left(\cos\varphi_S\cos2\varphi_P-\frac{1}{2}\sin\varphi_S\sin2\varphi_P\right)\bigg\}\ ,
\end{aligned}
\end{equation}
\begin{equation}
\label{gf0etaetap}
\begin{aligned}
\qquad\quad g_{f_0\eta\eta^\prime}=&\frac{\sin2\varphi_P}{2f_\pi}
\bigg\{(m^2_\eta\cos^2\varphi_P+m^2_{\eta^\prime}\sin^2\varphi_P-m^2_{a_0})
\left[\sin\varphi_S-\sqrt{2}\cos\varphi_S\left(2\frac{f_K}{f_\pi}-1\right)\right] \\[1ex]
&-(m^2_{\eta^\prime}-m^2_\eta)\left(\sin\varphi_S\cos2\varphi_P+\frac{1}{2}\cos\varphi_S\sin2\varphi_P\right)\bigg\}\ .
\end{aligned}
\end{equation}
\end{fleqn}
\end{widetext}
These couplings can be written in different equivalent forms;
here, the ones involving the $a_0$ mass and the pion decay 
constant have been chosen for the sake of clarity.
We can anticipate that taking into account the effects of scalar meson 
exchanges in an explicit way does not provide a noticeable improvement 
with respect to the chiral-loop prediction, except for the 
$\eta^\prime\to\eta\gamma\gamma$ case, where the $\sigma$ contribution turns 
out to be significant (cf.~Section \ref{results}).


\section{\label{results}Results and discussion}

In this section, use of the theoretical expressions developed
thus far is made to present quantitative results.
The decay widths for the processes of interest 
are calculated using the standard formula for the three-body 
decay \cite{PhysRevD.98.030001}, with the squared amplitude given by
\begin{equation}
\label{eqampsq}
\quad|\mathcal{A}|^{2}=|\mathcal{A}^{\rm{VMD}}|^{2}+|\mathcal{A}^{\rm{L\sigma M}}|^{2}+2{\rm{Re}}\mathcal{A}^{*\rm{VMD}}\mathcal{A}^{\rm{L\sigma M}}\,,
\end{equation}
where the vector ($\mathcal{A}^{\rm{VMD}}$) and scalar 
($\mathcal{A}^{\rm{L\sigma M}}$) exchange contributions have 
been presented in sections \ref{VMD} and \ref{LsM}, respectively. 
The last term in Eq.~(\ref{eqampsq}) represents the interference 
between the scalar and vector effects.

\begin{table}[b]
\centering
\begin{tabular}{llllll}
\hline
Decay &\qquad BR \cite{PhysRevD.98.030001} &\qquad $|g_{V\!P\gamma}|$ GeV$^{-1}$\\
\hline
$\rho^0\to\pi^{0}\gamma$			&\qquad $(4.7\pm0.6)\times10^{-4}$		&\qquad $0.22(1)$\\ 
$\rho^0\to\eta\gamma$				&\qquad $(3.00\pm0.21)\times10^{-4}$	&\qquad $0.48(2)$\\ 
$\eta^{\prime}\to\rho^0\gamma$	&\qquad $(28.9\pm0.5)\%$			&\qquad $0.40(1)$\\ 
$\omega\to\pi^{0}\gamma$		&\qquad $(8.40\pm0.22)\%$			&\qquad $0.70(1)$\\ 
$\omega\to\eta\gamma$			&\qquad $(4.5\pm0.4)\times10^{-4}$		&\qquad $0.135(6)$\\ 
$\eta^{\prime}\to\omega\gamma$	&\qquad $(2.62\pm0.13)\%$			&\qquad $0.127(4)$\\ 
$\phi\to\pi^{0}\gamma$			&\qquad $(1.30\pm0.05)\times10^{-3}$	&\qquad $0.041(1)$\\ 
$\phi\to\eta\gamma$				&\qquad $(1.303\pm0.025)\%$			&\qquad $0.2093(20)$\\ 
$\phi\to\eta^{\prime}\gamma$		&\qquad $(6.22\pm0.21)\times10^{-5}$	&\qquad $0.216(4)$\\ 
\hline                                           
\end{tabular}
\caption{PDG values for the branching ratios of the $V(P)\to P(V)\gamma$ 
transitions and the calculated $g_{V\!P\gamma}$ couplings directly from experiment 
(cf.~Eq.~(\ref{empirircalcoupling})).}
\label{gVPgammacouplingsempirical}
\end{table}

\begin{table*}
{\footnotesize{
  \centering
  \begin{tabular}{|c|l|c|c|c|c|c|c|c|c|c|}
\hline
    \multicolumn{1}{|c|}{Decay}& 
    \multicolumn{1}{c|}{Couplings}& 
    \multicolumn{1}{c|}{Chiral-loop}& 
    \multicolumn{1}{c|}{L$\sigma$M}& 
    \multicolumn{1}{c|}{VMD}&
    \multicolumn{1}{c|}{$\Gamma$}&
    \multicolumn{1}{c|}{BR$_{\rm th}$}&
    \multicolumn{1}{c|}{BR$_{\rm exp}$ \cite{PhysRevD.98.030001}}\\
    \hline
      \multirow{2}{*}{$\eta\to\pi^0\gamma\gamma$\,(eV)} & Empirical & $1.87\times 10^{-3}$ & $5.0\times 10^{-4}$ & 0.16(1) & 0.18(1) & $1.35(8)\times 10^{-4}$ & \multirow{2}{*}{$2.56(22)\times 10^{-4}$} \\ \cline{2-7}
& Model-based & $1.87\times 10^{-3}$ & $5.0\times 10^{-4}$ & 0.16(1) & 0.17(1) & $1.30(1)\times 10^{-4}$ & 
\\ 
   \hline
   \multirow{2}{*}{$\eta^{\prime}\to\pi^0\gamma\gamma$\,(keV)} & Empirical & $1.1\times 10^{-4}$ & $1.3\times 10^{-4}$ & 0.57(3)& 0.57(3) & $2.91(21)\times 10^{-3}$ & \multirow{2}{*}{$3.20(7)(23)\times10^{-3}$} \\ \cline{2-7}
& Model-based & $1.1\times 10^{-4}$ & $1.3\times 10^{-4}$ & 0.70(4) & 0.70(4) & $3.57(25)\times 10^{-3}$ & \\          
         \hline
         \multirow{2}{*}{$\eta^{\prime}\to\eta\gamma\gamma$\,(eV)} & Empirical & $1.4\times 10^{-2}$ &  3.29 & 21.2(1.2) & 23.0(1.2) & $1.17(8)\times10^{-4}$ & \multirow{2}{*}{$8.25(3.41)(0.72)\times 10^{-5}$} \\ \cline{2-7}
& Model-based & $1.4\times 10^{-2}$ & 3.29 & 19.1(1.0) & 20.9(1.0) & $1.07(7)\times 10^{-4}$ & 
\\ 
        \hline
  \end{tabular}
  \caption{Chiral-loop, L$\sigma$M and VMD predictions for the
  $\eta\to\pi^0\gamma\gamma$, $\eta^\prime\to\pi^0\gamma\gamma$ 
  and $\eta^\prime\to\eta\gamma\gamma$ decays with empirical and model-based 
  VMD couplings. 
  The total decay widths are calculated from the coherent sum of the L$\sigma$M 
  and VMD contributions. 
  \label{table1}
}}}
\end{table*}

For the numerical values of the masses and decay 
widths of the participating resonances, we use the most up-to-date 
experimental data from the PDG \cite{PhysRevD.98.030001}, 
whilst for the pion and kaon decay constants we employ 
$f_\pi=92.07$ MeV and $f_K=110.10$ MeV, repectively.
For the VMD couplings\footnote{Note that, for the L$\sigma$M couplings, 
\textit{i.e.}~$g_{\sigma\eta\eta^{\prime}}$ 
and $g_{f_0\eta\eta^{\prime}}$, the current experimental 
state-of-the-art does not allow obtaining the
associated numerical values directly from the empirical data. 
Therefore, one must resort to theoretical or phenomenological models 
to estimate them (cf.~Eqs.~(\ref{gsigmaetaetap}) and (\ref{gf0etaetap})). 
Likewise, the mixing angle in the scalar sector is fixed in our calculations 
to $\varphi_S=-8^\circ $ following Ref.~\cite{Escribano:2006mb}.} 
(cf.~Eq.~(\ref{AVMDetapi0})), we follow two different approaches: 
$i)$ The $g_{V\!P\gamma}$ are obtained directly from the experimental 
decay widths of the $V\to P\gamma$ and $P\to V\gamma$
($P=\pi^{0},\eta,\eta^{\prime}$ and $V=\rho^0,\omega,\phi$) 
radiative transitions \cite{PhysRevD.98.030001} 
by making use of 
\begin{equation}
\begin{aligned}
\Gamma_{V\to P\gamma}&=\frac{1}{3}\frac{g^{2}_{VP\gamma}}{32\pi}\left(\frac{m_{V}^{2}-m_{P}^{2}}{m_{V}}\right)^{3}\ ,\\
\Gamma_{P\to V\gamma}&=\frac{g^{2}_{VP\gamma}}{32\pi}\left(\frac{m_{P}^{2}-m_{V}^{2}}{m_{P}}\right)^{3}\ ,
\end{aligned}
\label{empirircalcoupling}
\end{equation}
and are summarised in Table \ref{gVPgammacouplingsempirical};
$ii)$ the phenomenological model from Ref.~\cite{Bramon:2000fr} is employed 
to parametrise the VMD couplings (cf.~Eq.~(\ref{eqcoups})), and, by performing 
an optimisation fit to the most up-to-date $V\!P\gamma$ 
experimental data \cite{PhysRevD.98.030001}, one can find 
preferred values for these 
parameters\footnote{Note that this phenomenological model, 
contrary to the one presented in Ref.~\cite{Escribano:2020jdy}, 
does not take into account isospin-violating effects and this is 
reflected in the quality of the fit, which is far from ideal, 
$\chi^2/\mathrm{d.o.f.}=5.3$. However, in this study we are working 
in the isospin limit and, therefore, this simplified version of 
the model suffices for our purposes. Should one have used 
more simplified models by setting, for example, $z_{\textrm{NS}}=1$ 
and $z_{\textrm{S}}=1$, or $z_{\textrm{NS}}=1$ 
and $z_{\textrm{S}}\overline{m}/m_s=1$, would lead to qualities of fits of 
$\chi^2/\mathrm{d.o.f.}=18.3$ and $\chi^2/\mathrm{d.o.f.}=110.4$, 
respectively, which are clearly not acceptable.}
\begin{equation}
\begin{gathered}
\begin{aligned}
g &= 0.70 \pm 0.01 \ \textrm{GeV}^{-1} \ ,	&	z_{\textrm{S}}\overline{m}/m_s = 0.65 \pm 0.01 \ ,\\
\phi_{P} &= (41.4 \pm 0.5)^\circ \ ,	&	\phi_{V} = (3.3 \pm 0.1)^\circ \ ,\\
\end{aligned}
\\
z_{\textrm{NS}} = 0.83 \pm 0.02 \ .
\end{gathered}
\label{eqfit}
\end{equation}
Hereafter, we refer to the former couplings as 
empirical and the later as model-based couplings. 

The numerical results obtained using both the empirical and model-based VMD 
couplings are summarised in Table \ref{table1}. There, we show the contributions 
from ChPT, the L$\sigma$M, which replaces ChPT when scalar meson 
poles are incorporated explicitly, and VMD. In addition, the theoretical decay widths
and corresponding branching ratios are presented, together with the associated
experimental values. Note that the 
quoted errors come from the uncertainties associated to the VMD couplings.
Using the empirical VMD couplings, one finds that, whilst our  
prediction for the $\eta\to\pi^{0}\gamma\gamma$ process, 
$\rm{BR}=1.35(8)\times10^{-4}$, is approximately a factor of two
smaller than the PDG reported value\footnote{Note that it is still compatible at the $\sim 5\sigma$ level with the experimental value though.}  \cite{PhysRevD.98.030001} 
$\rm{BR}=2.56(22)\times10^{-4}$, our theoretical estimates for the 
$\eta^{\prime}\to\pi^{0}\gamma\gamma$ and $\eta^{\prime}\to\eta\gamma\gamma$, 
$\rm{BR}=2.91(21)\times10^{-3}$ and $\rm{BR}=1.17(8)\times10^{-4}$, 
are consistent with the BESIII experimental measurements
\cite{Ablikim:2016tuo,Ablikim:2019wsb} $\rm{BR}=3.20(7)(23) \times10^{-3}$ 
and $\rm{BR}=8.25(3.41)(72)\times10^{-5}$, respectively.
Employing, instead, the model-based VMD couplings from 
Eq.~(\ref{eqcoups}) and making use of the fit values for the 
model parameters shown in Eq.~(\ref{eqfit}), we find that 
the branching ratio for the $\eta\to\pi^{0}\gamma\gamma$ decay, 
$\rm{BR}=1.30(8)\times10^{-4}$, is very much in line with that 
obtained using the empirical couplings, 
and approximately half the corresponding experimental 
value\footnote{Oset \textit{et al.}~considered additional 
contributions in Ref.~\cite{Oset:2008hp},
such as axial exchanges in the chiral loops and VMD loop contributions, 
where the associated amplitudes had been unitarised by making use of the 
Bethe-Salpenter equation for the resummation 
of the meson-meson scattering amplitudes,
as well as contributions from the three-meson axial anomaly;
all this allowed them to raise their prediction up to 
$\Gamma_{\eta\to\pi^0\gamma\gamma} = 0.33\pm 0.08$ eV.}.
Thus, our theoretical results for this reaction appear to be robust against small
variations of the VMD couplings. For the $\eta^{\prime}\to\pi^{0}\gamma\gamma$ 
and $\eta^{\prime}\to\eta\gamma\gamma$ processes, we obtain 
$\rm{BR}=3.57(25)\times10^{-3}$ and $\rm{BR}=1.07(8)\times10^{-4}$, which,
once again, are in agreement with the values reported by BESIII 
\cite{Ablikim:2016tuo,Ablikim:2019wsb}. The branching ratio for the later 
process turns out to be $\rm{BR}=1.11(8)\times 10^{-4}$ 
and $\rm{BR}=1.00(7)\times 10^{-4}$ for the empirical 
and model-based couplings using a Breit-Wigner propagator
for the $\sigma$ meson, where the pole parameters quoted in 
Ref.~\cite{PhysRevD.98.030001} have been utilised,
instead of the complete one-loop propagator. As can be seen, 
the use of either propagator provides very approximate results; 
any differences surface in the associated energy spectra.

\begin{figure*}
\centering
\begin{subfigure}[b]{0.49\textwidth}
	  \centering
        \includegraphics[height=0.24\textheight,width=0.9\textwidth,trim=7mm 0mm 0mm 0mm, clip=false]{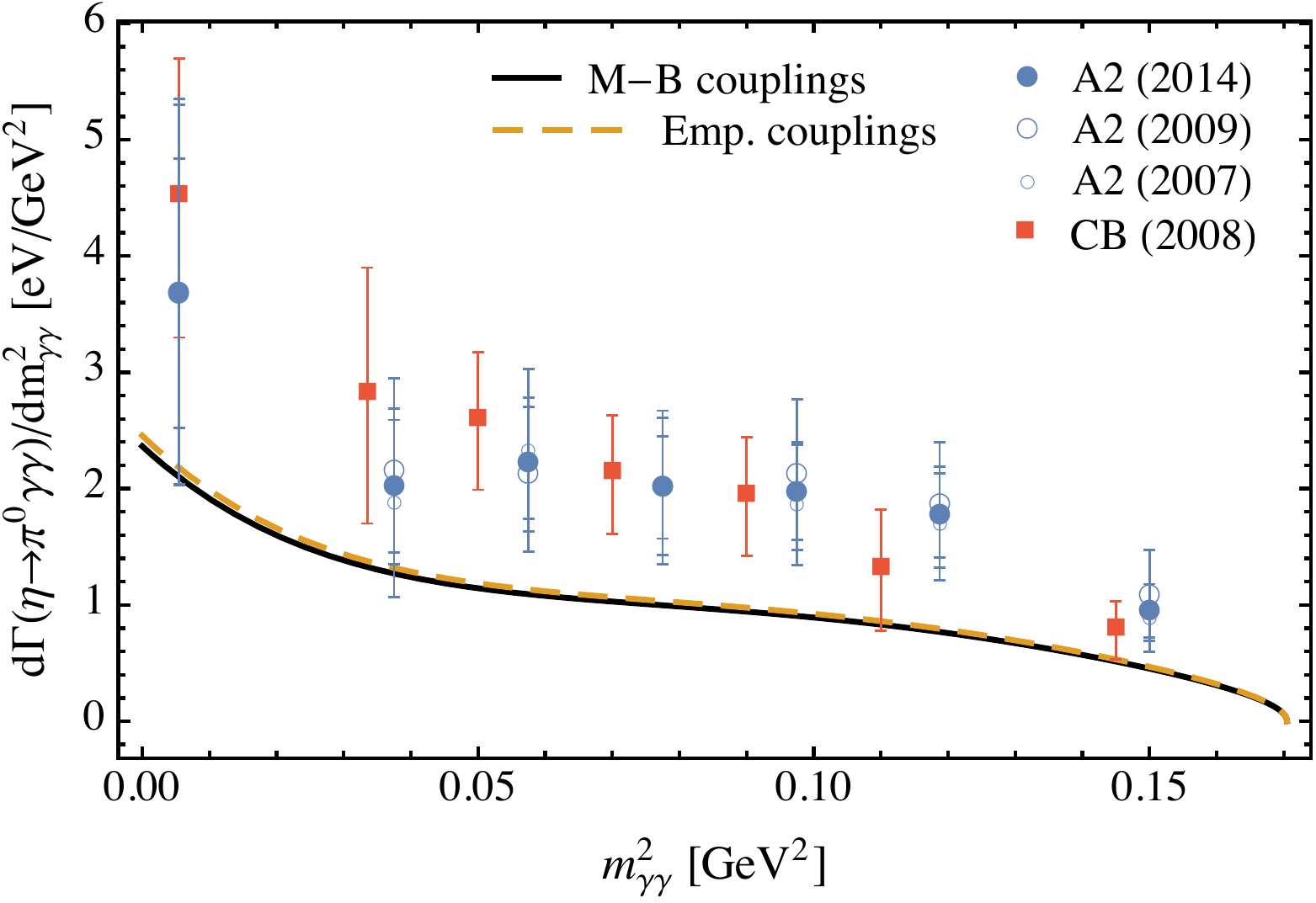}
        \caption{$\eta\to\pi^0\gamma\gamma$ decay.}
        \label{fig1_a}
\end{subfigure}
\begin{subfigure}[b]{0.49\textwidth}
	  \centering
        \includegraphics[height=0.235\textheight,width=0.92\textwidth,trim=0mm 0mm 0mm 0mm, clip=false]{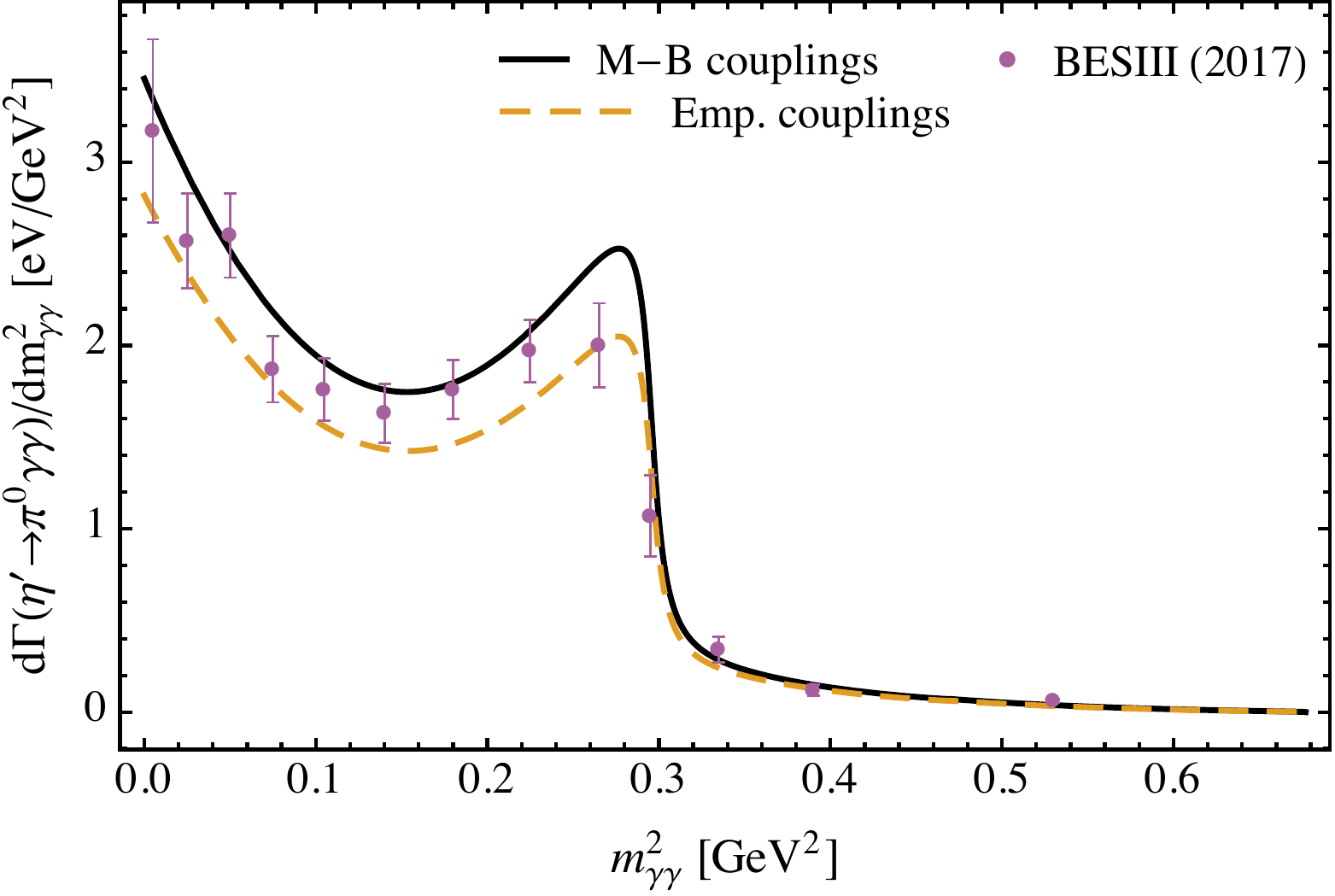}
        \caption{$\eta^{\prime}\to\pi^0\gamma\gamma$ decay.}
        \label{fig1_b}
\end{subfigure} \\[2\tabcolsep]
\caption{Comparison between the experimental diphoton energy spectra for 
the $\eta\to\pi^0\gamma\gamma$ and $\eta^{\prime}\to\pi^0\gamma\gamma$ 
and our theoretical predictions using the empirical and model-based
VMD couplings. The experimental data is taken from Ref.~\cite{Nefkens:2014zlt} 
(A2), Ref.~\cite{Prakhov:2008zz} (Crystal Ball) and Ref.~\cite{Ablikim:2016tuo} (BESIII).}\label{fig1}
\end{figure*}

\begin{figure*}
\begin{center}
\includegraphics[scale=0.7]{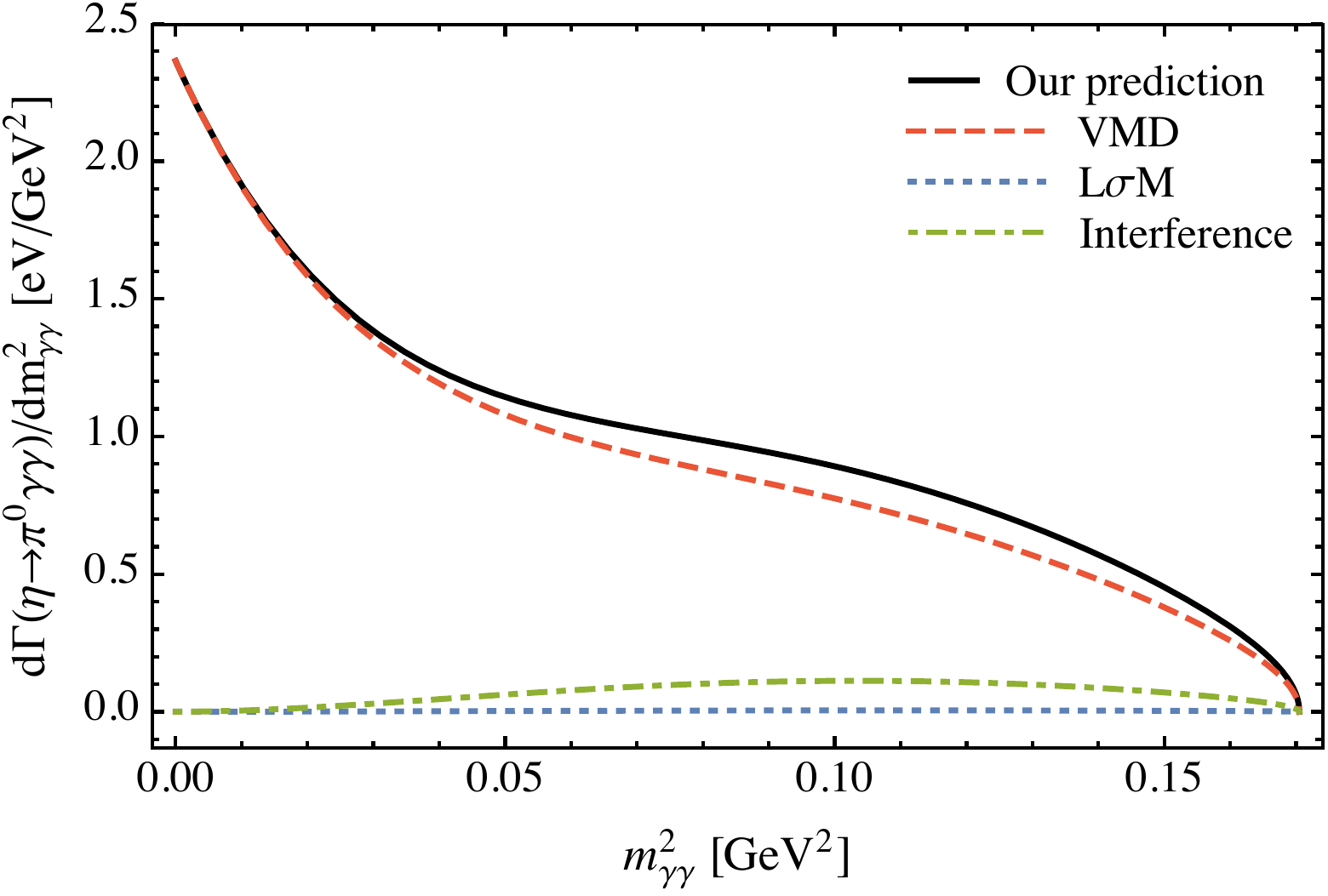}
\caption{Contributions to the $\eta\to\pi^{0}\gamma\gamma$ 
diphoton energy spectrum (solid black), using the model-based VMD 
couplings, from intermediate vector (dashed magenta) and scalar 
(dashed brown) meson exchanges, and their interference (dot-dashed cyan).}
\label{fig2}
\end{center}
\end{figure*} 

Our predictions for the diphoton energy spectra are compared 
with the corresponding experimental data in Fig.~\ref{fig1}. 
One can see from both plots that the shape of the spectra is captured 
well by our theoretical predictions. 
The spectrum of the $\eta\to\pi^{0}\gamma\gamma$ decay
(Fig.~\ref{fig1_a}) appears to present a normalisation offset\footnote{One 
could argue, though, that the experimental central values 
seem to lie further apart from our predictions for
decreasing $m_{\gamma\gamma}^2$, but this effect may be linked to 
the larger uncertainties associated to the measurements 
at low $m_{\gamma\gamma}^2$.}. 
Notwithstanding this, the exact same theoretical treatment shows very
good agreement between our predictions for the 
$\eta^{\prime}\to\pi^{0}\gamma\gamma$ spectrum, using either
set of VMD couplings, and experiment. 
In addition, the use of one set of couplings or the other makes 
little difference for the $\eta\to\pi^{0}\gamma\gamma$,
though, it appears that the model-based couplings capture slightly better the 
experimental data for the $\eta^\prime\to\pi^{0}\gamma\gamma$. For this reason, 
as well as due to its increased aesthetic appeal and the 
fact that it better underpins the power of the 
theoretical description, from this point onwards we will stick to using the 
model-based VMD couplings for any subsequent calculation.   

\begin{figure*}
\begin{center}
\includegraphics[scale=0.7]{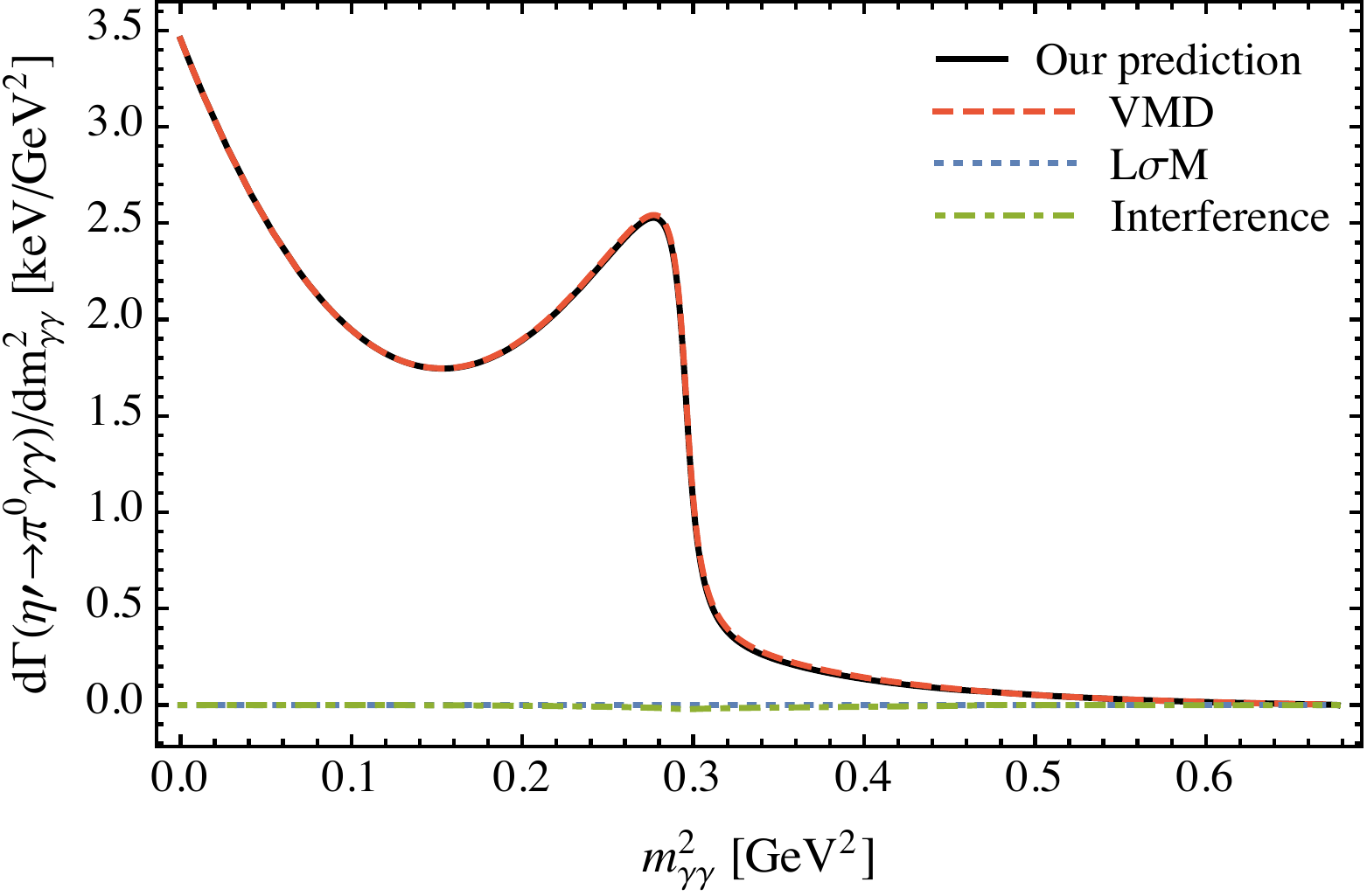}
\caption{Contributions to the $\eta^{\prime}\to\pi^{0}\gamma\gamma$ 
diphoton energy spectrum (solid black), using the model-based VMD 
couplings, from intermediate vector (dashed magenta) and scalar 
(dashed brown) meson exchanges, and their interference (dot-dashed cyan).}
\label{fig3}
\end{center}
\end{figure*} 

The different contributions to the diphoton energy spectrum for
the $\eta\to\pi^{0}\gamma\gamma$ decay are shown in Fig.~\ref{fig2}.
As it can be seen, the spectrum is dominated by the exchange of vector mesons, 
accounting for $93\%$, out of which, the weights for the $\rho^0$, $\omega$ 
and $\phi$ are $27\%$, $21\%$ and $0\%$, respectively;
the remaining $52\%$ comes from the interference between the three 
participating vector mesons.
The contribution of the scalar exchanges accounts for 
less than $1\%$, making it very 
difficult to isolate the effect of individual scalar mesons, even with the advent 
of more precise experimental data. The interference between the intermediate
scalar and vector exchanges is constructive and accounts for about $7\%$.
The contributions to the energy spectrum of the
$\eta^{\prime}\to\pi^{0}\gamma\gamma$ process are displayed in 
Fig.~\ref{fig3}. Once again, the exchange of vector mesons completely
dominate the spectrum contributing approximately with the $100.4\%$ to the 
total signal, whilst the effects of scalar meson 
exchanges and their interference with the formers are negligible with 
$0\%$ and $-0.4\%$ (destructive interference), respectively.
As well as this, the $\omega$ contribution prevails with the $78\%$ 
of the total VMD signal, whilst the $\rho^0$ and $\phi$ account for
the $5\%$ and $0\%$, respectively; the remaining $17\%$
comes from the interference between the vector resonances. 
\begin{figure*}
\begin{center}
\includegraphics[scale=0.7]{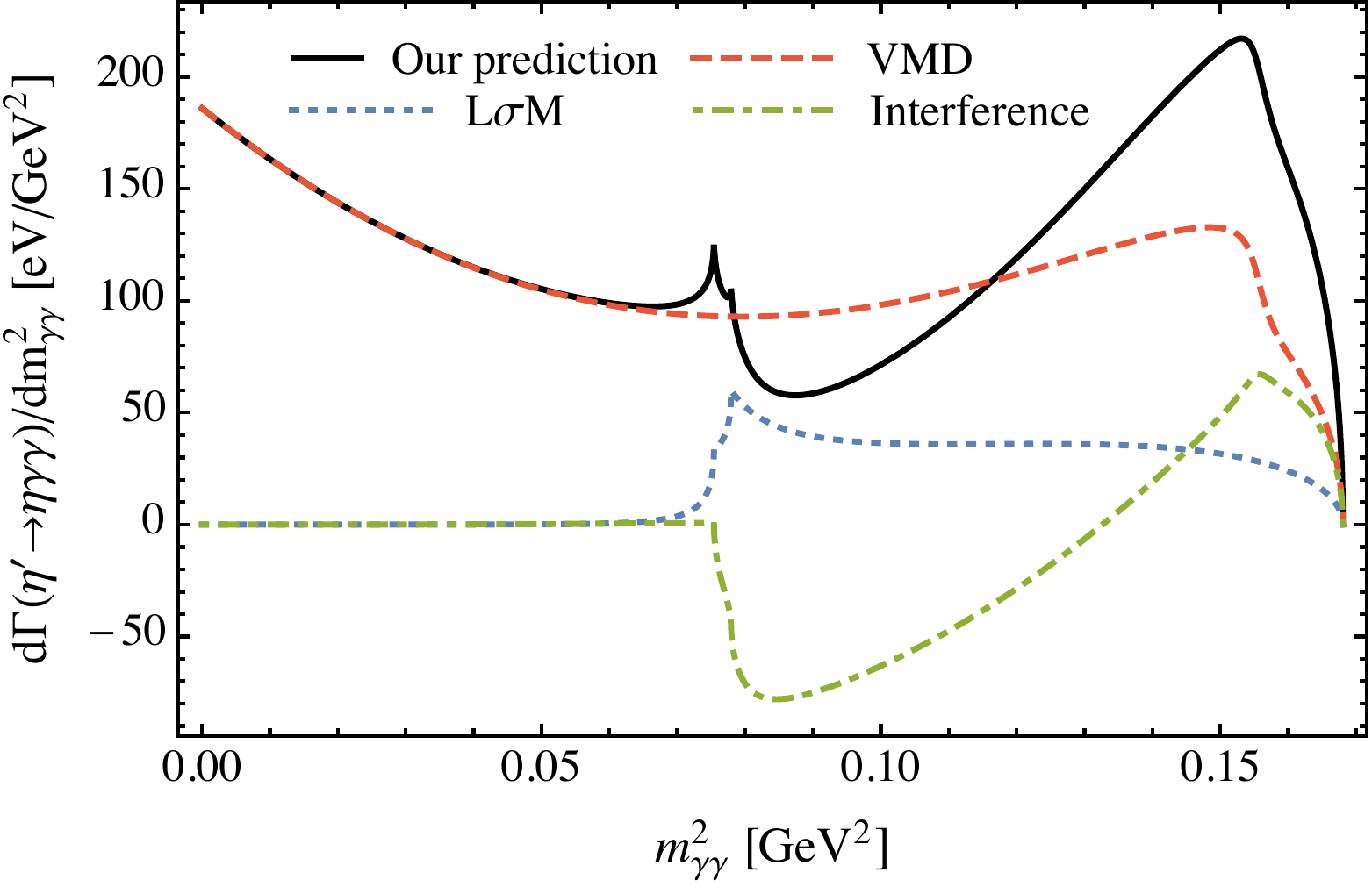}
\caption{Contributions to the $\eta^{\prime}\to\eta\gamma\gamma$ 
diphoton energy spectrum (solid black), using the model-based VMD 
couplings, from intermediate vector (dashed magenta) and scalar 
(dashed brown) meson exchanges, and their interference (dot-dashed cyan).}
\label{fig4}
\end{center}
\end{figure*} 
Finally, the different contributions to the $\eta^\prime\to\eta\gamma\gamma$ 
energy spectrum are presented in Fig.~\ref{fig4}. As expected,
the contribution to the total signal from the exchange of vector mesons 
predominates again with about the $91\%$, with the 
$\rho^0$, $\omega$ and $\phi$ accounting for
$59\%$, $15\%$ and $1\%$ of the VMD signal, respectively, and
the remaining $25\%$ being the result of their interference;
interestingly, the scalar meson effects turn out 
to be sizeable in this process, weighing approximately $16\%$, 
with the exchange of $\sigma$ mesons dominating the 
scalar signal\footnote{A possible improvement to our prediction
for the scalar meson contribution may be possible by considering
a more sophisticated scalar scattering amplitude 
${\cal A}_{\pi^{+}\pi^{-}\to\eta^{\prime}\eta}$ (cf.~Eq.~(\ref{ApippimetaetapChPTLsM})) 
as has successfully been done for the associated 
$\eta^{\prime}\to\eta\pi\pi$ decay process in Ref.\,\cite{Gonzalez-Solis:2018xnw}.}. 
The interference between the scalar
and vector mesons is destructive and accounts for around the 
$7\%$ and significantly influences the shape of the spectrum.
It is worth noting the effect of using the complete one-loop 
propagator for the $\sigma$ exchange which manifests at the 
$m_{\gamma\gamma}^2=0.078$ GeV peak and is associated to the 
$\pi^{+}\pi^{-}$ threshold. This peak is absent should the Breit-Wigner
propagator for the $\sigma$ exchange have been used.

Our inability to describe the total decay widths for the three 
$\eta^{(\prime)}\to\pi^{0}\gamma\gamma$ and 
$\eta^{\prime}\to\eta\gamma\gamma$ decay processes simultaneously 
within the same theoretical framework and values for the VMD couplings
is somewhat bothersome. The offset that appears to be 
affecting the diphoton energy spectrum of the first process, 
$\eta\to\pi^{0}\gamma\gamma$ (cf.~Fig.~\ref{fig1_a}), and consequently 
its integrated decay width, might be linked to a 
normalisation problem associated to the parameter $g$ in Eqs.~(\ref{eqcoups}) 
and (\ref{eqfit}). One could argue, though, that this parameter is fixed by the
$VP\gamma$ experimental data, which is measured nowadays to a high 
degree of accuracy and leads to satisfactory predictions for the other two processes,
\textit{i.e.}~$\eta^{\prime}\to\pi^{0}\gamma\gamma$ and 
$\eta^{\prime}\to\eta\gamma\gamma$, and, therefore, should not be changed.
Despite this, an attempt has been made to assess the \textit{preferred} 
value for the parameter $g$ by the experimental data available from 
Ref.~\cite{Nefkens:2014zlt} (A2), Ref.~\cite{Prakhov:2008zz} (Crystal Ball) and
Ref.~\cite{Ablikim:2016tuo} (BESIII) for the two 
$\eta^{(\prime)}\to\pi^{0}\gamma\gamma$ processes by performing a combined fit 
where $g$ is left as a free parameter. The resulting $g$ turns out to be 
roughly consistent with the one provided in Eq.~(\ref{eqfit})
and used in all our calculations, which is explained by the fact that the data from
BESIII contains significantly smaller uncertainties 
and, therefore, its statistical weight in the fit is greater.
Hence, we are led to consider whether this puzzle might be somehow 
highlighting the need for a more sophisticated theoretical treatment;
however, given the complexity associated to performing 
these experimental measurements and
the recent history of the $\eta\to\pi^{0}\gamma\gamma$ empirical  
data\footnote{For instance, in 1984 Alde 
\textit{et al.}~found $\mbox{BR}=7.2(1.4)\times 10^{-4}$
\cite{Alde:1984wj}, whilst more recent measurements appear to
indicate $\mbox{BR}=2.52(23)\times1 0^{-4}$ \cite{Nefkens:2014zlt} 
and $\mbox{BR}=2.21(24)(47)\times 10^{-4}$ \cite{Prakhov:2008zz}.}, 
one cannot rule out the possibility that this decreasing trend seen
over time in the measured values of the BR  
might persist should new and more precise measurements 
were available and eventually converge with our theoretical 
prediction, especially in light of our successful description of
the data from BESIII for the other two sister processes.


\section{\label{Conclusions}Conclusions}

In this work, we have presented a thorough theoretical analysis
of the doubly radiative decays $\eta^{(\prime)}\to\pi^0\gamma\gamma$ and 
$\eta\to\eta\gamma\gamma$, and provided theoretical 
results for their associated decay widths and diphoton energy spectra
in terms of intermediate scalar and vector meson exchange contributions
using the L$\sigma$M and VMD frameworks, respectively.


A complete set of theoretical expressions for the transition amplitudes from
Chiral Perturbation Theory, Vector Meson Dominance and the Linear Sigma 
Model have been given for the three decay processes. Some of these expressions 
constitute, to the best of our knowledge, the first predictions of this kind.
In addition, we have provided quantitative results by making use of numerical 
input from the PDG \cite{PhysRevD.98.030001}. In particular, for 
the estimation of the VMD coupling constants, $g_{V\!P\gamma}$, 
two different paths have been followed 
whereby they have been either extracted directly from the experimental  
$V(P)\to P(V)\gamma$ decay widths or from a phenomenological 
quark-based model and a fit to experimental data. A summary of the predicted
decay widths, theoretical branching ratios and contributions to the total signals for
the three doubly radiative decays $\eta\to\pi^0\gamma\gamma$ and 
$\eta^\prime\to \pi^0(\eta)\gamma\gamma$ is shown in Table \ref{table1},
and a discussion of the results obtained and how they compare 
to available experimental data has been carried out. 
As well as this, the invariant mass spectra associated to these processes 
are shown in Figs.~\ref{fig2}, \ref{fig3} and \ref{fig4}, respectively, 
using the model-based VMD couplings.
It is worth highlighting that, whilst vector meson exchanges vastly dominate 
over the scalar contributions for the $\eta^{(\prime)}\to\pi^{0}\gamma\gamma$
decays, we find that, for the $\eta^{\prime}\to\eta\gamma\gamma$, the scalar meson 
effects turn out to be substantial, specially that of the $\sigma$ meson, 
and this represents an opportunity for learning details about this still poorly 
understood scalar state. In particular, we look forward to the release of 
the energy spectrum data for the $\eta^{\prime}\to\eta\gamma\gamma$ process
by the BESIII collaboration to assess the robustness of our theoretical approach.

Interestingly, our predictions for the $\eta\to\pi^{0}\gamma\gamma$ are found to be 
approximately a factor of two smaller than the experimental measurements, 
whereas our theoretical predictions for the $\eta^{\prime}\to\pi^{0}\gamma\gamma$ 
and $\eta^{\prime}\to\eta\gamma\gamma$ are in good agreement with 
recent measurements performed by BESIII. It appears that it is not possible to
reconcile our predictions for the three processes with their corresponding 
experimental counterparts simultaneously using the same underlying theoretical 
framework and values for the coupling constants. This puzzle might be pointing
towards potential limitations of our theoretical treatment or, perhaps, the need
for more precise measurements for the $\eta\to\pi^{0}\gamma\gamma$ decay,
as our approach seems to be capable of successfully predicting the experimental
data for the other two processes 
without the need for manual adjustment of the numerical input.

As a final remark, we would very much like to encourage experimental groups 
to measure these decays once again to confirm whether our predictions 
are correct or a more refined theoretical description is required.\\


\begin{acknowledgements}

The authors would like to thank Feng-Kun Guo for a careful reading 
of the first manuscript, Andrzej Kupsc for insisting on the relevance 
of this work in view of the recent BESIII data for the
$\eta^{\prime}\to\pi^{0}\gamma\gamma$ and 
$\eta^{\prime}\to\eta\gamma\gamma$ decay processes, 
and Manel L\'{o}pez Meli\`{a} for 
pointing out a typo in a previous version of the manuscript.
The work of R.~Escribano and E.~Royo is supported by the 
Secretaria d'Universitats i Recerca del Departament d'Empresa i 
Coneixement de la Generalitat de Catalunya under the grant 2017SGR1069, 
by the Ministerio de Econom\'{i}a, Industria y Competitividad under 
the grant FPA2017-86989-P, and from the Centro de Excelencia Severo Ochoa 
under the grant SEV-2016-0588. This project has received funding from 
the European Union's Horizon 2020 research and innovation 
programme under grant agreement No.~824093.
S.~Gonz\`{a}lez-Sol\'{i}s received financial support from the 
CAS President's International Fellowship Initiative for Young 
International Scientists (Grant No.\,2017PM0031 and 2018DM0034), 
by the Sino-German Collaborative Research Center \textquotedblleft 
Symmetries and the Emergence of Structure in QCD\textquotedblright\,
(NSFC Grant No.\,11621131001, DFG Grant No.\,TRR110), 
by NSFC (Grant No.\,11747601), and by the National Science 
Foundation (Grant No. PHY-2013184). 

\end{acknowledgements}


\appendix

\section{\label{propagators}Complete one-loop propagators}
The complete one-loop propagators for the $\sigma$, $f_0$ and $a_0$ 
scalar resonances are defined as follows
\begin{equation}
\label{propagator}
D(s)=s-m_R^2+{\rm Re}\Pi(s)-{\rm Re}\Pi(m_R^2)+ i{\rm Im}\Pi(s)\ ,
\end{equation}
where $m_R$ is the renormalised mass of the scalar meson and
$\Pi(s)$ is the one-particle irreducible two-point function.
${\rm Re}\Pi(m_R^2)$ is introduced to regularise the divergent behaviour 
of $\Pi(s)$. The propagator so defined is well behaved when 
a threshold is approached from below,
thus, improving the usual Breit-Wigner prescription, which is not particularly 
suited for spinless resonances
(see Ref.~\cite{Escribano:2002iv} for details).

The real and imaginary parts of $\Pi(s)$ for the $\sigma$ in the
first Riemann sheet\footnote{We follow the convention from Ref.~\cite{Bhattacharya:1991gr} for the definition of the first Riemann
sheet of the complex square root and complex logarithm functions.} 
can be written as $(R(s)\equiv {\rm Re}\Pi(s),I(s)\equiv {\rm Im}\Pi(s))$
\begin{widetext}
\begin{fleqn}
\begin{equation}
\begin{aligned}
\label{Rf0}
\qquad\qquad\qquad R(s)=&\frac{g_{\sigma\pi\pi}^2}{16\pi^2}
\bigg[2-\beta_\pi\log\left(\frac{1+\beta_\pi}{1-\beta_\pi}\right)\theta_\pi
-2\bar\beta_\pi\arctan\left(\frac{1}{\bar\beta_\pi}\right)\bar\theta_\pi\bigg] \\[2pt]
&+\frac{g_{\sigma K\bar K}^2}{16\pi^2}\bigg[2-\beta_K\log\left(\frac{1+\beta_K}{1-\beta_K}\right)\theta_K
-2\bar\beta_K\arctan\left(\frac{1}{\bar\beta_K}\right)\bar\theta_K\bigg]\ , 
\end{aligned} 
\end{equation}\\[-6ex]
\begin{equation}
\begin{aligned}
\label{If0}
\qquad\qquad\qquad I(s)=-\frac{g_{\sigma\pi\pi}^2}{16\pi}\beta_\pi\theta_\pi-\frac{g_{\sigma K\bar K}^2}{16\pi}\beta_K\theta_K\ ,
\end{aligned}
\end{equation}
\end{fleqn}
\end{widetext}
where $\beta_i=\sqrt{1-4m_i^2/s}$ for $i=(\pi, K)$,
$\bar\beta_i=\sqrt{4m_i^2/s-1}$, $\theta_i=\theta(s-4m_i^2)$, and 
$\bar\theta_i=\theta(4m_i^2-s)$.
The couplings of the $\sigma$ to pions and kaons are written in the 
isospin limit\footnote{In our analysis, 
we work in the isospin limit and, therefore, the mass difference
between $K^0$ and $K^+$ is not taken into account for the $K{\bar K}$ threshold.}; thus, $g_{\sigma\pi\pi}^2=\frac{3}{2}g_{\sigma\pi^+\pi^-}^2=\frac{3}{2}\Big(\frac{m_{\pi}^2-m_{\sigma}^2}{f_{\pi}}\cos{\varphi_S}\Big)^2$ and 
$g_{\sigma K\bar K}^2=2g_{\sigma K^+K^-}^2=\frac{1}{2}\Big[\frac{m_{K}^2-m_{\sigma}^2}{f_{K}}(\cos{\varphi_S}-\sqrt{2}\sin{\varphi_S})\Big]^2$. The renormalised mass of the
$\sigma$ meson for the calculations is fixed to $m_{\sigma}=498$ MeV
\footnote{This value is obtained by solving the corresponding
pole equation $D(s_P)=0$, with $s_P=m_P^2-im_P\Gamma_P$, in the 
second Riemann sheet and ensuring that the pole mass and width 
are in accordance with the experimental data.}.
For the $f_0$ exchange, the real and imaginary parts of the two-point 
function in the first Riemann sheet are
\begin{widetext}
\begin{fleqn}
\begin{equation}
\begin{aligned}
\label{Rf0}
\qquad\qquad\qquad R(s)=&\frac{g_{f_0\pi\pi}^2}{16\pi^2}
\bigg[2-\beta_\pi\log\left(\frac{1+\beta_\pi}{1-\beta_\pi}\right)\theta_\pi
-2\bar\beta_\pi\arctan\left(\frac{1}{\bar\beta_\pi}\right)\bar\theta_\pi\bigg] \\[2pt]
&+\frac{g_{f_0 K\bar K}^2}{16\pi^2}\bigg[2-\beta_K\log\left(\frac{1+\beta_K}{1-\beta_K}\right)\theta_K
-2\bar\beta_K\arctan\left(\frac{1}{\bar\beta_K}\right)\bar\theta_K\bigg]\ , 
\end{aligned} 
\end{equation}\\[-6ex]
\begin{equation}
\begin{aligned}
\label{If0}
\qquad\qquad\qquad I(s)=-\frac{g_{f_0\pi\pi}^2}{16\pi}\beta_\pi\theta_\pi-\frac{g_{f_0 K\bar K}^2}{16\pi}\beta_K\theta_K\ ,
\end{aligned}
\end{equation}
\end{fleqn}
\end{widetext}
where $\beta_i$, $\bar\beta_i$, $\theta_i$ and 
$\bar\theta_i$ are defined as before.
Once again, the couplings of the $f_0$ to pions and kaons are written in the 
isospin limit; accordingly, $g_{f_0\pi\pi}^2=\frac{3}{2}g_{f_0\pi^+\pi^-}^2=\frac{3}{2}\Big(\frac{m_{\pi}^2-m_{f_0}^2}{f_{\pi}}\sin{\varphi_S}\Big)^2$ and 
$g_{f_0 K\bar K}^2=2g_{f_0 K^+K^-}^2=\frac{1}{2}\Big[\frac{m_{K}^2-m_{f_0}^2}{f_{K}}(\sin{\varphi_S}+\sqrt{2}\cos{\varphi_S})\Big]^2$.
The renormalised mass of the
$f_0$ meson for the calculations is fixed to $m_{f_0}=990$ MeV. 
%
Finally, the real and imaginary parts of $\Pi (s)$ for the $a_0$ in the first Riemann 
sheet are
\begin{widetext}
\begin{fleqn}
\begin{equation}
\begin{aligned}
\label{Ra0}
\qquad\qquad\qquad R(s)=&\frac{g_{a_0 K\bar K}^2}{16\pi^2}
\bigg[2-\beta_K\log\left(\frac{1+\beta_K}{1-\beta_K}\right)\theta_K
-2\bar\beta_K\arctan\left(\frac{1}{\bar\beta_K}\right)\bar\theta_K\bigg] \\[2pt]
&+\frac{g_{a_0\pi\eta}^2}{16\pi^2}
\bigg[2-\frac{m^2_\eta-m^2_\pi}{s}\log\left(\frac{m_\eta}{m_\pi}\right)
-\beta^+_{\pi\eta}\beta^-_{\pi\eta}
\log\left(\frac{\beta^-_{\pi\eta}+\beta^+_{\pi\eta}}{\beta^-_{\pi\eta}-\beta^+_{\pi\eta}}\right)
\theta_{\pi\eta} \\[2pt]
&-2\bar\beta^+_{\pi\eta}\beta^-_{\pi\eta}\arctan\left(\frac{\beta^-_{\pi\eta}}{\bar\beta^+_{\pi\eta}}\right)
\bar\theta_{\pi\eta}
+\bar\beta^+_{\pi\eta}\bar\beta^-_{\pi\eta}
\log\left(\frac{\bar\beta^+_{\pi\eta}+\bar\beta^-_{\pi\eta}}{\bar\beta^+_{\pi\eta}-\bar\beta^-_{\pi\eta}}\right)
\bar{\bar\theta}_{\pi\eta}\bigg]\ ,
\end{aligned} 
\end{equation}\\[-6ex]
\begin{equation}
\begin{aligned}
\label{Ia0}
\qquad\qquad\qquad I(s)=-\frac{g_{a_0 K\bar K}^2}{16\pi}\beta_K\theta_K
      -\frac{g_{a_0\pi\eta}^2}{16\pi}\beta^+_{\pi\eta}\beta^-_{\pi\eta}\theta_{\pi\eta}\ ,
\end{aligned}
\end{equation}
\end{fleqn}
\end{widetext}
where $\beta^\pm_{\pi\eta}=\sqrt{1-(m_\pi\pm m_{\eta})^2/s}$,
$\bar\beta^\pm_{\pi\eta}=\sqrt{(m_\pi\pm m_{\eta})^2/s-1}$,
$\theta_{\pi\eta}=\theta[s-(m_\pi+m_{\eta})^2]$,
$\bar\theta_{\pi\eta}=\theta[s-(m_\pi-m_{\eta})^2]\times\theta[(m_\pi+m_{\eta})^2-s]$, and 
$\bar{\bar\theta}_{\pi\eta}=\theta[(m_\pi-m_{\eta})^2-s]$.
The couplings of the $a_0$ to kaons are also written in the isospin limit; 
hence, $g_{a_0 K\bar K}^2=2g_{a_0 K^+K^-}^2=\frac{1}{2}\Big(\frac{m_{K}^2-m_{a_0}^2}{f_{K}}\Big)^2$ and $g_{a_0 \pi \eta}^2=\Big(\frac{m_{\eta}^2-m_{a_0}^2}{f_{\pi}}\cos{\varphi_P}\Big)^2$. For the calculations, 
the renormalised mass of the
$a_0$ meson is fixed to $m_{a_0}=980$ MeV. 


\end{document}